\documentclass[manuscript]{aastex}

\shorttitle{Can Nanoflares Heat the} \shortauthors{Tajfirouze \& Safari}

\begin{document}


\title{Can a Nanoflare Model of EUV Irradiances Describe the
Heating of the Solar Corona?}
\author{E. Tajfirouze,  H. Safari}
\affil{Dept. of Physics, Zanjan University, P. O. Box 45195-313,
Zanjan, Iran }
%

\begin{abstract}
  Nanoflares, the basic unit of impulsive energy release may produce much of the solar background emission.
 Extrapolation of the energy frequency distribution of observed microflares, which follows a power law to lower energies
 can give an estimation of the importance of nanoflares for heating the solar corona.
If the power law index is greater than 2, then the nanoflare
contribution is dominant.

 We model time series of extreme ultraviolet
 emission radiance, as random flares with a power law exponent of the flare  event distribution.
  The model is based on three key parameters, the flare
 rate, the flare duration and the power law exponent of the flare
  intensity frequency distribution.
  We use this model to simulate emission line radiance detected in 171 \AA, observed by STEREO/EUVI and SDO/AIA.
 The Observed light curves are matched  with simulated
light curves using an Artificial Neural Network and
  parameter values are determined across regions of active region, quiet sun, and coronal hole.
The damping rate of nanoflares is compared with radiative
losses cooling time. The effect of background emission, data cadence, and network
sensitivity on the key parameters of model is studied.

  Most of the observed light curves have a power law exponent, $\alpha$,  greater than the critical value 2.
  At these sites nanoflare heating could be significant.
\end{abstract}
\section{Introduction}

The mechanism of forming nanoflares is the dissipation of current
sheets arisen from tangential discontinuities in the continuously
evolving corona (Levine 1974, Parker 1988). Parker presumes that
the change in the magnetic field across the current sheet, $\Delta
B$, is critical for the onset. When the strength $|\Delta{B}|$ of
the discontinuity exceeds some threshold, there is a runaway
dynamical instability leading to an explosive reconnection phase.
This is similar to the sand pile model that has been used to
explain the comparison is also made between results of different
methods. Avalanches of magnetic reconnection
 (Lu \& Hamilton 1991), pointed out that the magnetic
field of the corona is in a state of self organized criticality.
To determine whether nanoflares are the main source of heat input
to the corona or not, it is necessary to measure the
energy frequency distribution of the smallest observed flares.

Hudson (1991) pointed out that the flare occurrence follows a
power law distribution such as, $dN\sim E^{-\alpha}dE$, in which
 $dN$ is the number of flares per energy interval $E$ and $E+dE$  (Lu
\& Hamilton (1991) obtained this distribution for complex systems
which are in a self organized critical state). The power law
index, $\alpha$, is a critical value for determining whether more
weight is given to small scale events (nanoflares) or larger ones
(flares). Determination of the power law index of the flare
frequency distribution is a scientific challenge. Many authors
have attempted to find this index.
 Benz \& Krucker (1998), using Yohkoh and SoHO,
shown that the power law index at the  microflare frequency
distribution is 2.5. Parnell \& Jupp (2000), used TRACE
observations and concluded that the energy of nanoflares is
insufficient to heat the quiet sun corona. Aschwanden \& Parnell
(2002) also get some other power law distribution by combining
 scaling law and fractal geometry for different observations.
Aschwanden \& Charbonneau (2002), gathered the statistics of
solar flares, microflares, and nanoflares and concluded that, the
power law index  falls bellow the critical value. An extended
review on the simulation and observational results is given by (
Klimchuk 2006 and Klimchuk et al. 2009).

  {To determine the contribution of small scale events (nanoflares) on the solar coronal heating,
 some applicable models has been investigated (e. g., Vekstein \& Katsukawa 2000, Sakamoto et al. 2009,
 Terzo et al. 2011). Here,} we use a model to simulate the observed
Extreme Ultra Violet (EUV) emitted radiation from STEREO
and SDO which has been applied successfully to the UV radiance
fluctuation in the quiet Sun (Pauluhn
 \& Solanki 2007) and later to SUMER observations of the corona in
 an active region (Bazarghan et al. 2008).

This paper is organized as follow: STEREO/EUVI and SDO/AIA data
analysis are described in Section \ref{data}. The
nanoflare model is treated in Section \ref{model}.  A
kind of Artificial Neural Networks (Probabilistic Neural Network)
is briefly discussed in Section 4. In sections
\ref{method} and \ref{result} the method and results are
presented.  The conclusions are given in Section
\ref{conc}.

\section{STEREO/EUVI and SDO/AIA data analysis}\label{data}
STEREO (Solar TErrestrial RElation Observatory) is the mission in
NASA`s Solar Terrestrial Probes program (STP) which has
 been launched on October 2006 probes solar active phenomena and dynamics.
 It contains two identical observatories,
 one ahead (A) and the other behind (B) of the orbit of the
Earth. Imaging the same events from two different locations
enables us to have three dimensional images
 (\url{http://stereo.gsfc.nasa.gov/}).
\begin{figure}[ht]
\centering
\includegraphics[width=\linewidth]{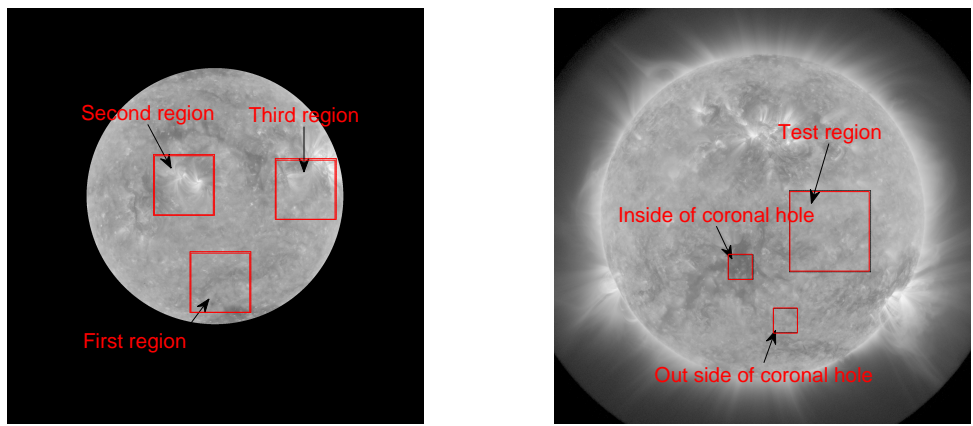}
       \caption{ (Left) The full disk  STEREO/EUVI 171 \AA  ~image of the Sun and the three selected regions
       outlined with red boxes.
       This image has been recorded on 13 June 2007. (Right) The full disk
        SDO/AIA 171 \AA~ image of the Sun recorded on 22 August 2010.}
            \label{fig1}
\end{figure}

 The Extreme
Ultra Violet Imager (EUVI) telescope is a part of SECCHI
instrument suite which is sensitive at 171 \AA,
195 \AA, 284 \AA, and 304 \AA. In the present work we restrict
ourselves to 171 \AA ~images with time cadence of 2.5 minutes
(W\"{u}elser et al. 2004). The spatial pixel size of both A and B
image is 1.6 arc second.

The SDO (Solar Dynamics Observatory) is designed to
simultaneously study the solar atmosphere at
small scales of space and time and in many wavelengths
(\url{http://sdo.gsfc.nasa.gov/\\mission/instruments.php}).  It
was launched on 11 February  2010.

The SDO/AIA provides full-disk imaging of the Sun in several UV
and EUV bands with a pixel size of about 0.5 arcsecond and a
cadence of 10 seconds.

Here, we take: (a) An STEREO/EUVI 171 \AA~ data set that has been
recorded on 13 June 2007 with a cadence of 2.5 min, (b)
SDO/AIA 171 \AA ~data set that has been recorded on 22 August ,
2010 with a cadence of 90 seconds.

The following steps are taken for data analysis:
\begin{description}
 \item[-]  Calibration of the recorded images using secchi$\_{\rm prep}$.pro on Solar
 Soft for EUVI data. The AIA level 1.5 data in the series, aia\_test.synoptic2 are
 used. These are binned data with a pixel size of 2 arcsecond and a cadence of 90s.
  \item[-]
To remove the solar differential rotation we use the Solar Soft
routine derot$\_{\rm map}$.pro.
\item[-]  EUVI and AIA provide full disk images.
We select three regions as shown in Figure \ref{fig1}  for EUVI
data and two regions for the AIA data. It is attempted
to select regions which cover active regions, quiet Sun
 and coronal hole.
Consider the first selected region of EUVI images. We partitioned
it into smaller $3\times3$ pixels regions. Averaging on
intensities of these smaller regions for $573 $ success images,
we obtain a light curve. The same process was followed for other
small remained areas of selected region. In similar manner some
other light curves were obtained too, while partitioning the
selected region into $5\times5$ and $9\times9$ pixels regions.
The above process was followed for other selected regions (on EUVI
and AIA) too. We note that the light curves have 573 and 939 time
steps for EUVI and AIA respectively.  We fed each light curves
(observed time series) to the network to determine the nanoflare
model parameters  (Section \ref{result}).
\end{description}

\section{Nanoflare Model}\label{model}
 Assuming that the coronal EUV emission is caused by many random flares with flare radiance, following
 a power law frequency distribution we take a model which
 consists of a time series of random kicks according to the presumption that flaring is
 intrinsically a stochastic process, each is given by a rising time $\tau_r$ and a damping
time $\tau_d$ (Pauluhn \& Solanki 2007, Safari et al. 2007, and Bazarghan et al. 2008).
This model has five free parameters. The maximum and minimum flare amplitude, $y_{max}$
 and $y_{min}$, the power law exponent $\alpha$,
the damping time $\tau_d$ and the flare rate $p_f$. These are
sufficient to produce our simulated EUV emission light curves.

Pauluhn \& Solanki (2007) have shown that for a large number of
independent random flares, the distribution of normalized
radiance in quiet sun follows a lognormal. The same is true  for
active regions as pointed out by Safari et al. (2007) and
Bazarghan et al.(2008).

\begin{figure}[ht]
\centering
\includegraphics[width=\linewidth]{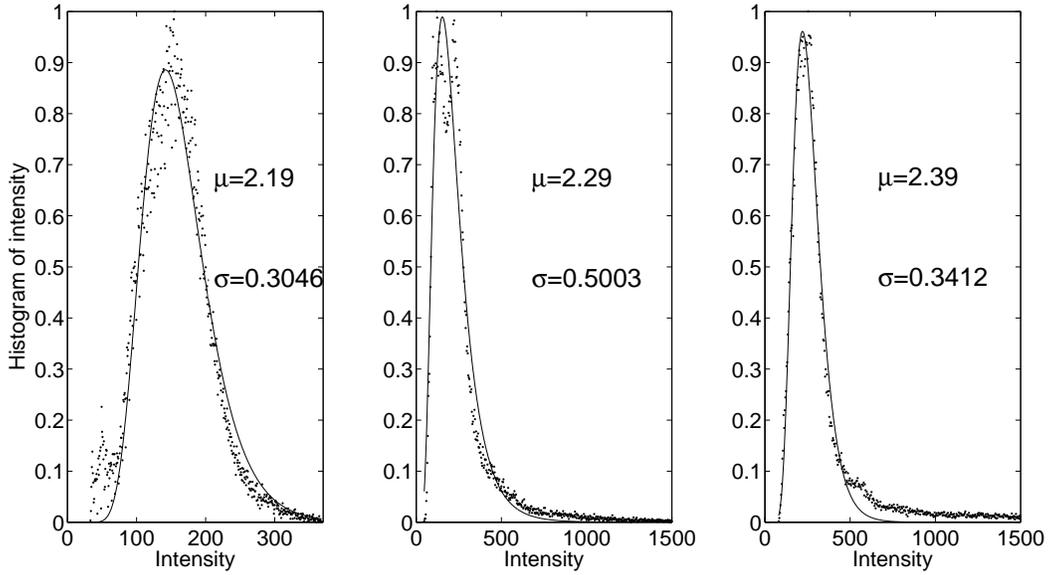}
       \caption{ Left to right, the histogram of
        intensity for the first, second, and third selected regions and their own lognormal
        fits.
 }
            \label{fig2}
\end{figure}

 In the following, we focus on the
distributions of three selected  regions shown in  Figure
\ref{fig1}.  The intensity distributions and the lognormal fits
are shown in Figure \ref{fig2}. The lognormal function is given by
\begin{equation}
f(x,\mu,\sigma)=\frac{1}{x\sqrt{\pi\sigma}}\exp\left(-\left(\frac{\log
x-\mu}{\sqrt{2}\sigma}\right)^2\right), \end{equation} where,
$x$, $\mu$, and $\sigma$ are the radiance,  scale parameter, and
shape parameter, respectively.

 The good fits  give us confidence in applying a
stochastic flare model. The key parameters (power law exponents,
damping rates and flare rates) of small scale events (small
eruptions) are not identifiable directly from observations.
Pauluhn \& Solanki (2007) and Safari et al. (2007),
 estimated that the damping rates and flare rates could be deduced by comparing light
 curves with simulated time series.
They have used the shape parameters of lognormal fits and wavelet analysis.

 {It is interesting to compare our model and it's
parameters to that of represented in recent works.
 Vekstein \& Katsukawa (2000) assume that the energy distribution of the nanoflares has
a power law spectrum, bounded between some minimum and maximum
energy release. They suggest that: each nanoflares create a
filament with cross-section area which has been then divided into
small grids; nanoflares are generated randomly inside these grids
with a finite occurrence rate;  and evolves with two stage of
cooling process (conduction and radiation heat). The heat
conduction and radiation are derived according to scaling low (e.
g., Rosner et al. 1978, Cargill 1993, Cargill \& Klimchuk 1997).
Sakamoto et al. (2009) have added the pressure balance (between
internal gas and external magnetic field of elemental filaments)
to this model and they analysed the SXT and TRACE intensities and
intensities fluctuations. Terzo et al. (2011) introduce a
constant heating as an intrinsically flat light curves and
imposed the Poisson noise to simulate the light curves. They
concluded that, the noises could not simulate the light curves
observed Hinode/SXT. Performing a Monte Carlo simulation they
perturb the flat light curves with a sequence of random segments
of exponential decays linked together. The parameters of their
model are the e-folding time, the interval between two successive
perturbation and the amplitude. Finally they found some simulated
light curves which is in best match with the observed data.
 }

\begin{figure}[ht]
\centering
\includegraphics[width=\linewidth]{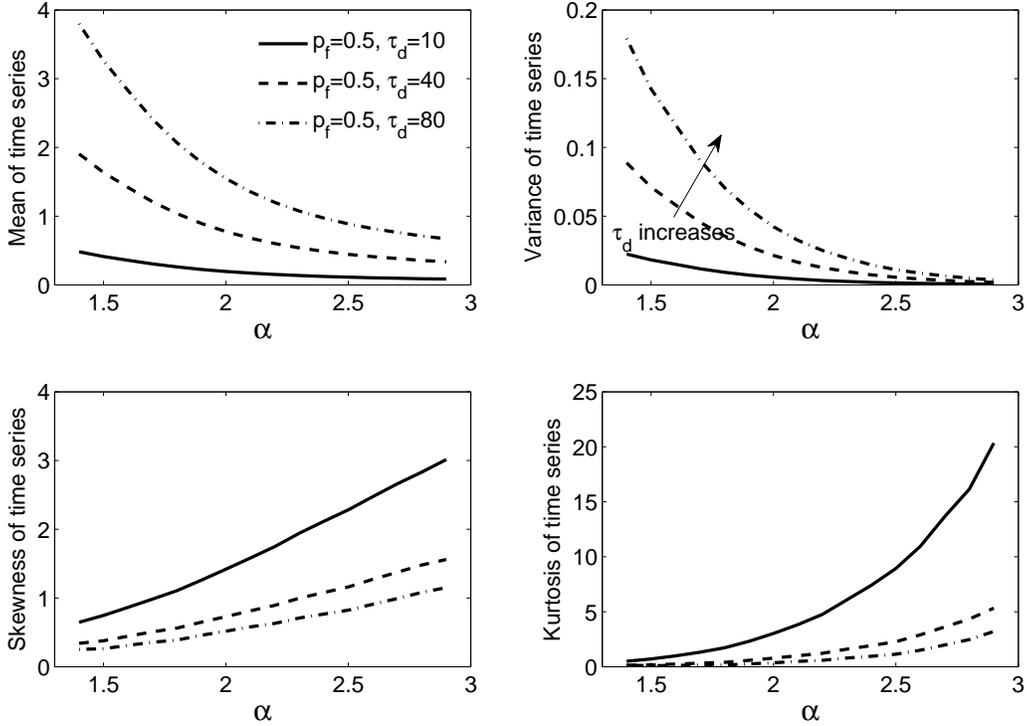}
       \caption{Mean, variance, skewness, and kurtosis of
       the light curves versus $\alpha$ for different $\tau_d$. }
            \label{fig3}
\end{figure}

\begin{figure}[ht]
\centering
\includegraphics[width=\linewidth]{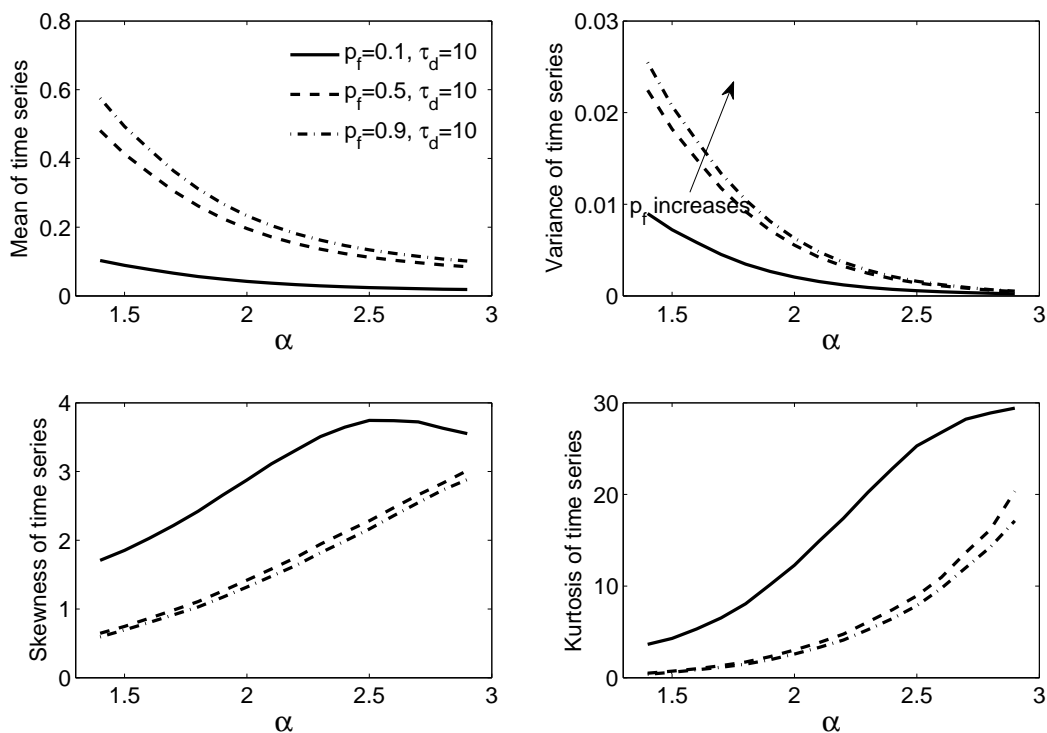}
       \caption{Mean, variance, skewness, and kurtosis of
       the time series versus $\alpha$ for different $p_f$.    }
            \label{fig4}
\end{figure}

\begin{figure}[ht]
\centering
\includegraphics[width=\linewidth]{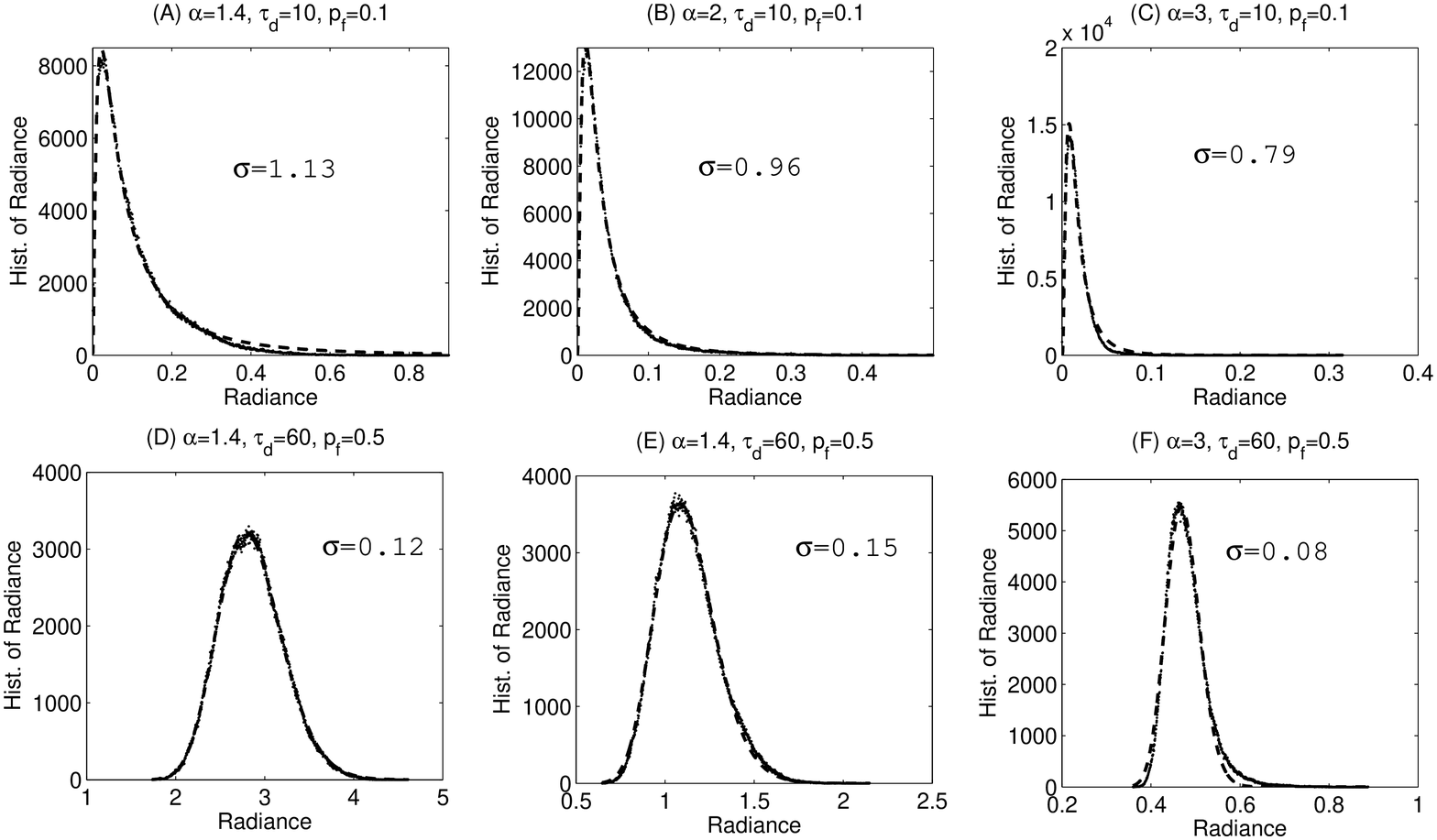}
       \caption{ Distribution functions of the simulated time series (dotted lines)
        and the best fit lognormal functions (dashed lines). The shape parameter
        varied considerably if $p_f\tau_d$ changed.}
            \label{fig5}
\end{figure}

 One may ask what happens to shape of the time series
 if the key parameters change. The light curves for flare models with different $\alpha$,
  $p_f$, and $\tau_d$ parameters are shown( Bazarghan et al. 2008, Figure 4).
 There is a physical picture of how the parameters affect the light curves.
 The mean, variance, skewness, and kurtosis  of the time series
 are shown in Figures \ref{fig3} and \ref{fig4} versus $\alpha$.
   Both the mean and variance of the time series decrease, as $\alpha$ increases.
   Expectedly, the more the $\alpha$ (corresponds to more small events) the less the intensities  become.
 With a fixed $\alpha$ and $p_f$ as $\tau_d$ increases, the mean and variance increases accordingly. This also
 occurs when $\alpha$ and $\tau_d$ are fixed and $p_f$ changes. The
higher moments (skewness  and kurtosis)  of the  time series
increase with increasing $\alpha$, $\tau_d$, and $p_f$.
Due to lognormal shape of the distributions, both
skewness and kurtosis values are  positive (Figure \ref{fig5}).
 A positive value of skewness signifies a distribution with an
asymmetric tail extending out towards larger radiance.  {In
general, the geometric median ($\exp(\mu)$) is less than the
geometric mean ($\exp(\mu+\sigma^2/2)$). This is agree with Terzo
et al. (2011). They have shown that, the distributions of
intensity fluctuations  taken from individual pixels, multi-pixel
subregions, or the entire active region as observed by Hinode/SXT
are asymmetric.  }

 We knew that, the shape parameter is inversely in proportion
 to $\sqrt{\tau{p_f}}$ (Pauluhn \& Solanki 2007) and that the lognormal shape
 parameter, $\sigma$, slightly changes with $\alpha$.  Although,
  the sharpness of the distribution indicates that for a higher
 $\alpha$, the distribution shape tends to be sharpener as illustrated by the kurtosis increase (Figure
 \ref{fig5}).

Bazarghan et al. (2008)  used Artificial Neural Networks  (ANN) to
compare  SUMER/SoHO observational time series with simulated time
series. In this paper, a similar method is employed to determine
the three key parameters of SECCHI/EUVI and SDO/AIA time series.
 The main advantages of the
ANN method is that it enables us to obtain quantitative
values for all parameters including $\alpha$,
 with which  Safari et al.(2007) had problem in
analysis.


\section{Probabilistic Neural Network}\label{Artificial Neural Network}
\begin{figure}[ht]
\centering
\includegraphics[width=\linewidth]{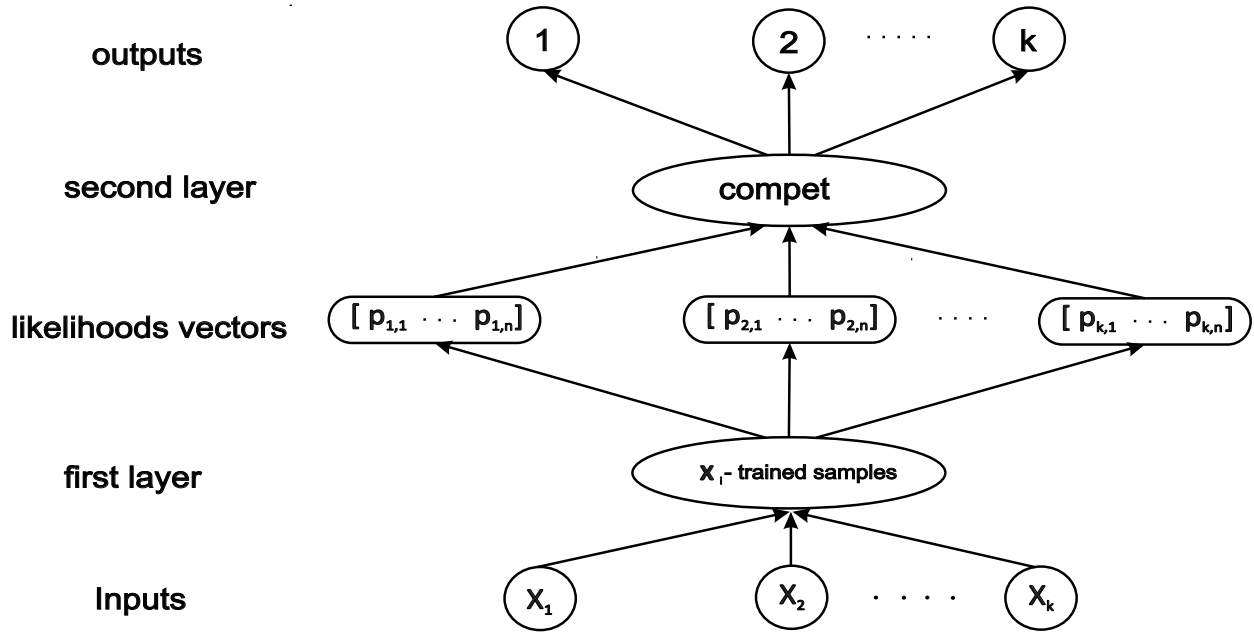}
       \caption{ The architecture of Probabilistic Neural Network}
            \label{fig6}
\end{figure}

We employed a kind of  Artificial Neural Networks, called
Probabilistic Neural Network (PNNs Specht 1988, 1990) which is
suitable for classification . A classification problem is defined
with a set of inputs $P$ and targets $T$. PNN is a kind of
supervised network. It means that the learning process of the
network takes place with an initially specified set of inputs and
targets, called trained samples. If we assume that the input
vectors contain $k$ different classes,
 then every target vector would contains $k$ elements. One of them is 1,
 which corresponds to its own class and the others are zero.
   The PNN has two layers( Figure \ref{fig6}). When an input vector is fed to the network,
the first layer calculates the distance between the input vector
and the trained samples. So, it provides a vector the elements of
which define the distance between the input and trained samples.
  The second layer produces a vector which contains the
probabilities using the output of the first layer. Finally the
competitive transfer function of the second layer chooses the
maximum likelihood value within the probabilities vector and
produces the output $1$  for it and the output zero for the
others. Using this process, the network matches the input vector
to one of the existing $k$ classes which has the maximum
likelihood value.

 We ran the commands of the mentioned process in  the
Neural Network Toolbox (Wasserman 1993).

\section{Methods}\label{method}
We employed the following steps:
\begin{description}
 \item[1-]  We ran the simulation code to generate
the light curves. The power law exponent ranges between
1.4 $\leq\alpha\leq$4.4 in steps of 0.1; the duration time
ranges between 2$\leq\tau_d\leq80$ in steps of 1,
 the flare rate falls between 0.1$\leq  p_f<$1 in steps of 0.1.
    {A total amount of 22041
 light curves for each combination of $\alpha$, $\tau_d$, and $p_f$  are obtained.
  For series with labels 1 and 7800 the set
had
 ($\alpha=1.4$, $\tau_d=2$, $p_f=0.1$) and
  ($\alpha=2.4$, $\tau_d=8$, $p_f=0.6$), respectively. The larger the label's value is, the greater
  the value of the power law index becomes.
}
\item[2-] We fed the simulated light curves (as produced in
previous steps) to the network.  {Now, the network was tested to
see whether it has been well trained or not?
 A set of simulated light curves (trained samples) was selected (manually) to feed the network
  to enable it to recognize and classify. If the network was able to recognize then it would be trained.
 Since our model has been based on random number, we reproduced some light curves
 and fed them to the network for more accurate testing. We note
 that due to the network which uses a special function, the training error of the network is zero.}
 \item[3-]  { The input weights of first layer of the network is a matrix, elements of which
are the trained samples couples (input and targets). Let us
consider an observed light curve which the set of it's key
parameters are not known. Once the unknown light curve (test
sample) is fed to the network as an input vector, it's distance
from the 22041 trained vectors is calculated by the first layer.
So, it would be determined how close the input is to the vector
of trained set. Consequently, for an input vector closest to
trained one, the existent transfer function of the first layer
produces 1. If the input vector has the same distance from two or
more trained samples, then the output includes two or more 1. The
second layer weights are set to a matrix, elements of which are
the target vectors. Each vector has only a 1 in the row
associated with that particular class of input, and 0's
elsewhere. The output of first layer is multiplied by the matrix
(target) and the product will be sent to the transfer function of
second layer which produces 1 for the greatest existent value and
0's elsewhere. Using these process, the network corresponds the
input vector to one of the 22041 existent
  classes because that class had the maximum probability of being
  correct. The output of the classifier
  includes a number which  labeled each classes, each corresponds to a set of key
  parameters. Now, the total observed light curves are fed to
  the classifier (network) as the test data set.}

\end{description}

\begin{figure}[ht]
\centering
\includegraphics[width=\linewidth]{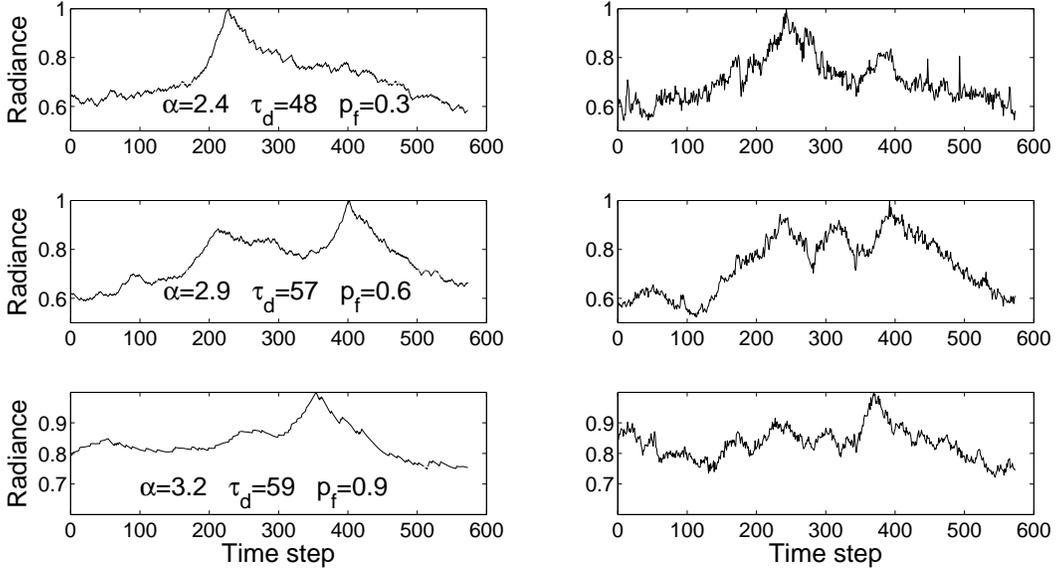}
       \caption{ Left panel:  The simulated light curves for three set of key
parameters are shown as legends with $\tau_r/\tau_d=0.3$ and
$y_{\rm max}/y_{\rm min}=100$. Right Panel:  The observed light
curves from EUVI 171 \AA, for the first  selected regions while
averaging on intensities of the smaller regions ($3\times3$,
$5\times5$, and $9\times9$ pixels from above to bottom,
respectively). Both left and right compared to each other using
Artificial Neural Network.}
            \label{fig7}
\end{figure}
 \begin{figure}[ht]
\centering
\includegraphics[width=\linewidth]{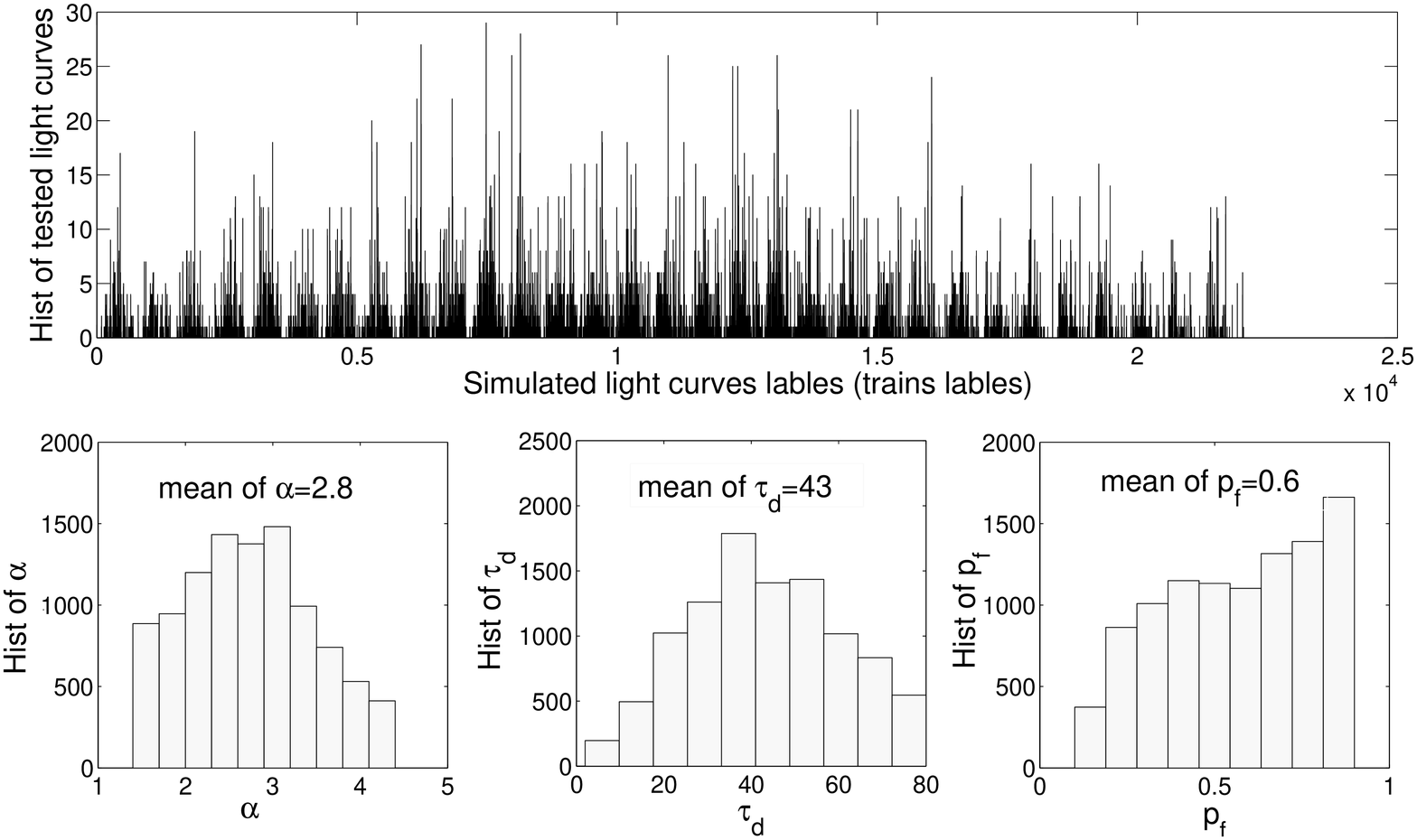}
       \caption{ Above panel:  The frequency (histogram) of output labels (simulated light curves labels)
        for the observed light curves ( First region of STEREO/EUVI image) constructed by averaging on intensities
        of the smaller region ($3\times3$ pixels). Bottom panel (from left to right):
The histograms of $\alpha$, $\tau_d$, and $p_f$, respectively.}
            \label{fig8}
\end{figure}
\begin{figure}[ht]
\centering
\includegraphics[width=\linewidth]{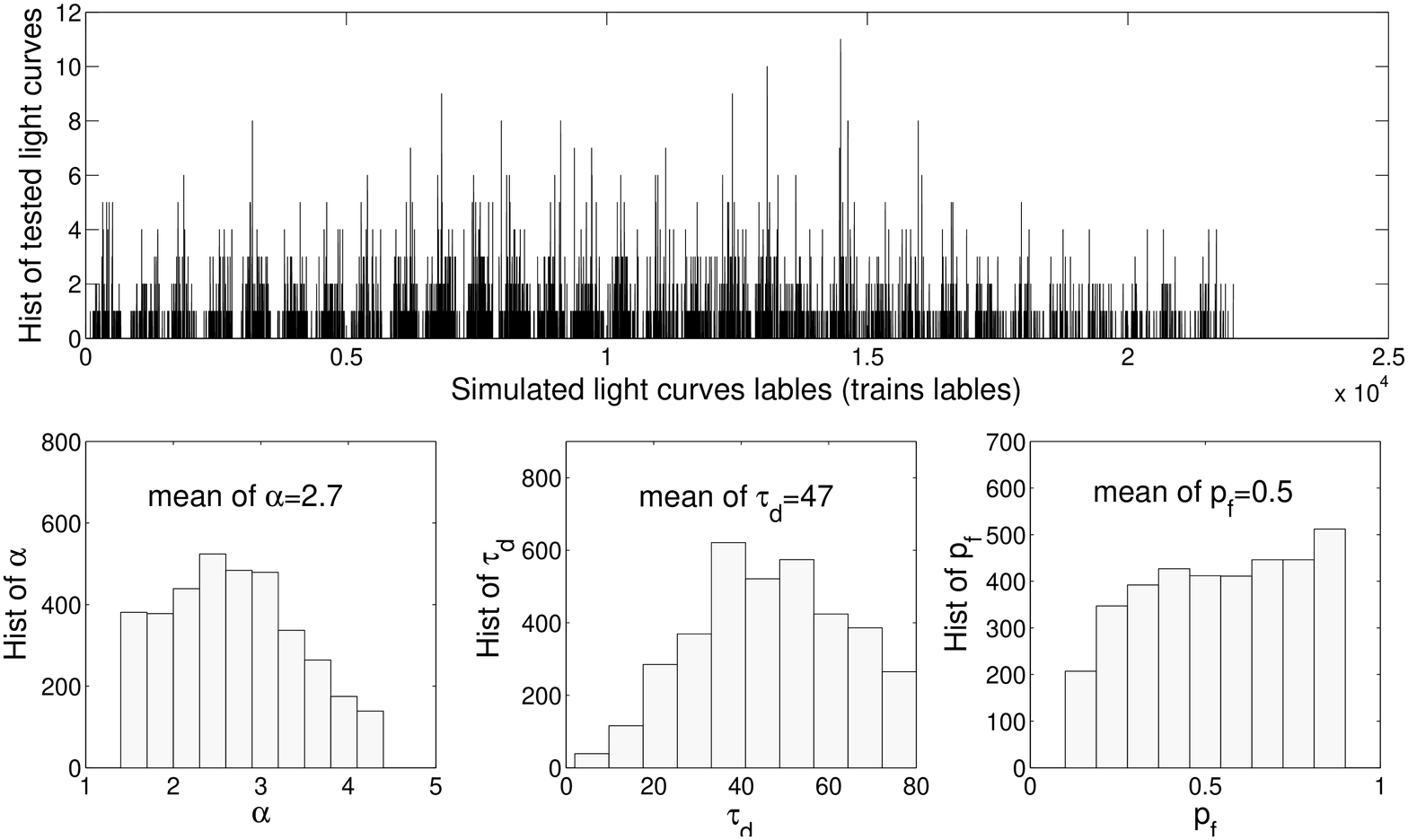}
       \caption{Above panel:  The frequency (histogram) of output labels (simulated light curves labels)
        for the observed light curves ( First region of STEREO/EUVI image) constructed by averaging on intensities
        of the smaller region ($5\times5$ pixels). Bottom panel (from left to right):
The histograms of $\alpha$, $\tau_d$, and $p_f$, respectively. }
            \label{fig9}
\end{figure}
\begin{figure}[ht]
\centering
\includegraphics[width=\linewidth]{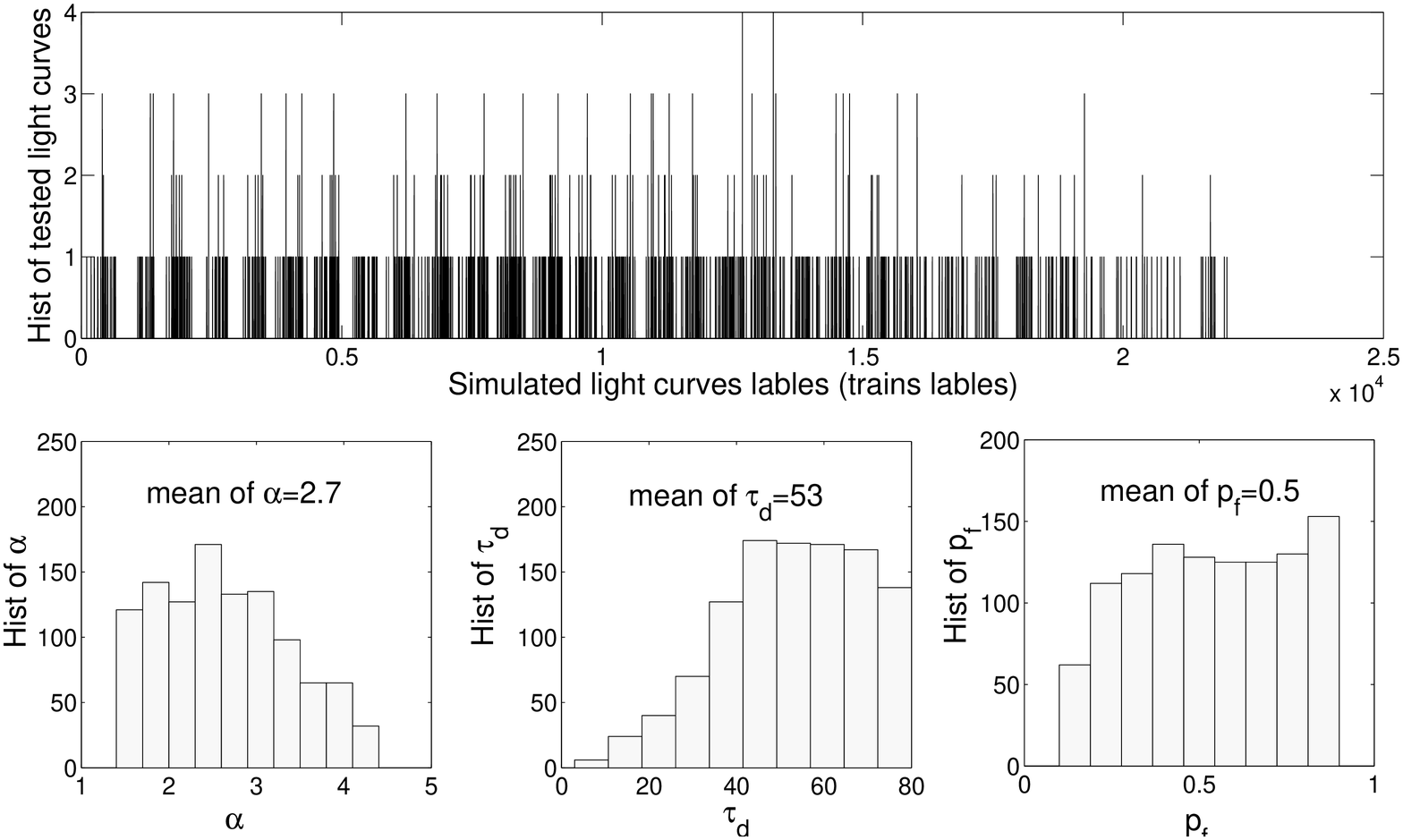}
       \caption{Above panel:  The frequency (histogram) of output labels (simulated light curves labels)
        for the observed light curves ( First region of STEREO/EUVI image) constructed by averaging on intensities
        of the smaller region ($9\times9$ pixels). Bottom panel (from left to right):
The histograms of $\alpha$, $\tau_d$, and $p_f$, respectively.}
            \label{fig10}
\end{figure}
\section{Results}\label{result}
\subsection{Results for STEREO/EUVI}
For each of the three selected regions (300$\times$300
pixels area) as marked on Figures \ref{fig1}, the light curves for
average intensities were made. This gives, a set of 10000 light
curves while averaging on the intensities of the smaller
$3\times3$ pixels regions, 3600 light curves while averaging on
the intensities of smaller $5\times5$ pixels regions, and 1089
light curves while the selected smaller regions are found  to be
$9\times9$ pixels. These light curves were fed to the network.
The network labeled each light curve with a set of individual key
parameters which corresponds to a simulated light curve. The
output result of the network is shown in Figures
\ref{fig7}-\ref{fig10}.

In Figure \ref{fig7}, the observed light curves are
compared with their matched simulated light curves  found by the
network.
 We see that, the background
radiance of the simulated light curves are remarkably close to the
observed ones, which was a problem in previous studies of SUMER
data (Bazarghan et al. 2008, see Figures 6 therein). As wee see,
the observed light curves are noisier than the simulated one. One
may wonder what would be the effect of the light curves' noise on
the network's output?  Using wavelet automatic de-noising (with
different threshold and standard deviations) the noise of
observed light curves is removed. In this case, the network's
output did not change.

The frequency of the observed light curves (tested
samples) versus simulated light curves (trained samples)
 is shown in the upper panels in Figures \ref{fig8},
\ref{fig9}, and \ref{fig10}. To do this, the histogram of output
labels is calculated. We found that, these labels ranged from 1
to 22041, which corresponded to simulated light curves for which
the key parameters range from the set of
$(\alpha=1.4,\tau_d=2,p_f=0.1)$ to
$(\alpha=4.4,\tau_d=80,p_f=0.9)$, as was stated in the previous
section. In the bottom panels, the frequencies (histogram) of
power law index, damping time, and flare rate, are extracted as
well.

 As we see in Figures  \ref{fig8}, \ref{fig9}, and \ref{fig10}
 the network's output for $\alpha$ are concentrated on the range between 2-3.
 The mean power law index, $\alpha$, is 2.8, 2.7 and
 2.6 for intensity averaged over  $3\times3$, $5\times5$ and $9\times9$
 pixels, respectively.   {There are the similar values for the standard deviations,
 $\sigma\approx0.7$.
 As was expected, the power law index falls as the
 the dimension of pixel's average intensity increases.}
  This confirms that,
 larger areas involve larger flare events.
We note that the $\tau_d$ values increase with binning. For
example,
 the average $\tau_d$ for $3\times3$ pixels is 43 and for $9\times9$ pixels it is 53.
  The $\tau_d$ seems to depend on the event size.
 This suggests that the larger areas as was expected have greater background.

 The response of the STEREO/EUVI 171 \AA~ as a function
of plasma temperature is within the range  $\log
T_e\approx5.1-6.7$ (W\"{u}lser et al. 2004). If we suppose the
plasma cooling through this narrow band filter is dominated by
radiative losses process,
 then the cooling time, $\tau_{\rm cool}\approx\tau_{rad}$, would be ranged
from a few seconds to an hour (see e. g., Aschwanden 2000, 2004).
 Using the mean dimensionless $\tau_d=43$ and multiplying it by cadence of 2.5 min
  (for STEREO/EUVI, 13 June 2007), we obtain a value of 2 hours.
  This is in good agreement with  previous results (Aschwanden 2004).

 Similar results are obtained for the two other selected regions. For the clarification of
 the background effects, we point at the SDO/AIA data in the following section.

\subsection{Results for SDO/AIA}
To explain the background effect on the three key
parameters, $\alpha$,  $\tau_d$, and $p_f$, we used SDO/AIA 171
\AA~ data sets taken on  22 August 2010. To do this, a region
from inside and a region outside  a coronal hole were selected
(Figure \ref{fig1}, right panel). The time series with average
intensity of $3\times3$ pixels were constructed and fed to the
network.

\begin{figure}[ht]
\centering
\includegraphics[width=\linewidth]{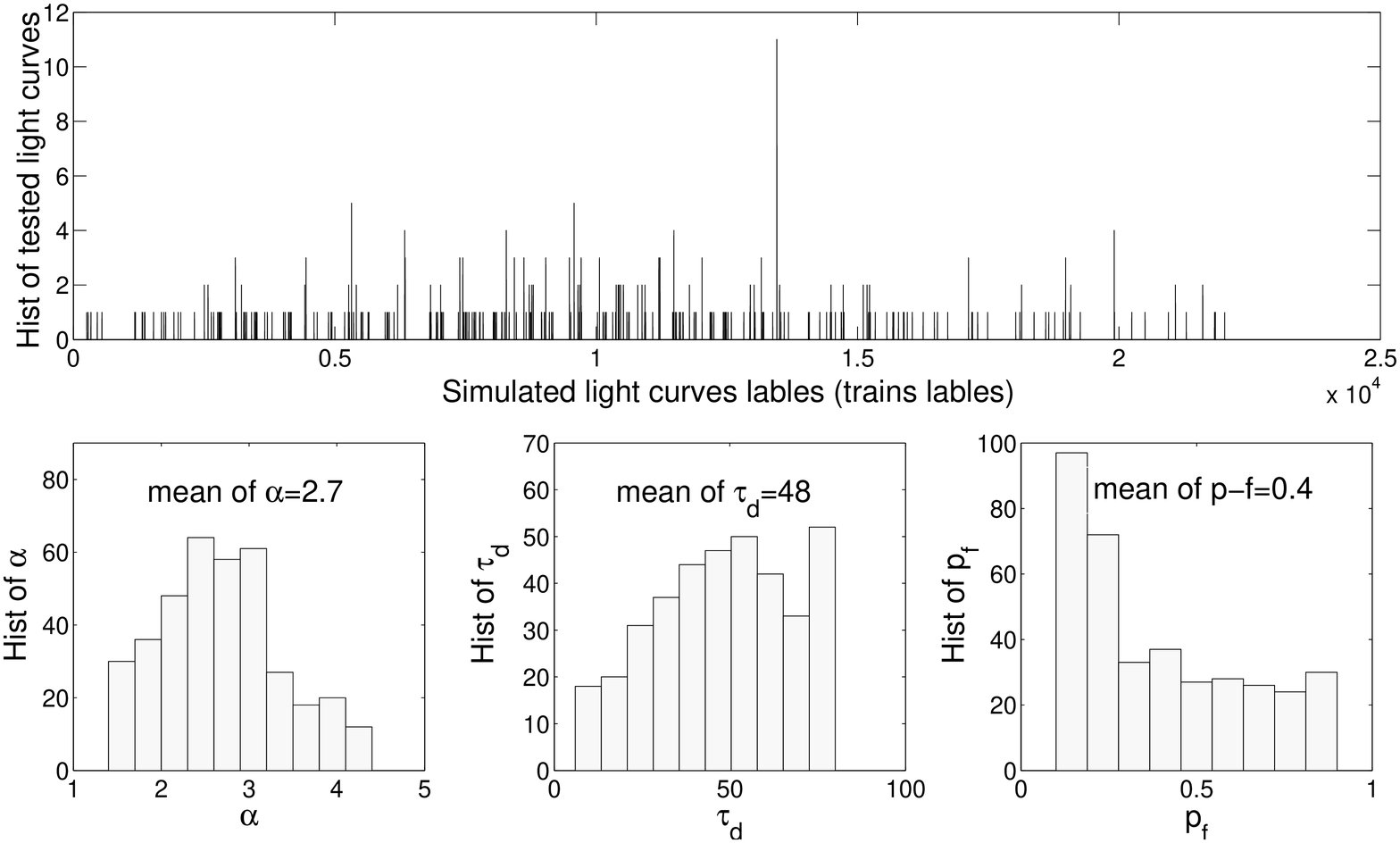}
       \caption{Above panel:  The frequency (histogram) of output labels (simulated light curves labels)
        for the observed light curves (Inside of coronal hole of SDO/AIA image)  constructed by averaging on intensities
        of the smaller region ($3\times3$ pixels). Bottom panel (from left to right):
The histograms of $\alpha$, $\tau_d$, and $p_f$, respectively. }
            \label{fig11}
\end{figure}

\begin{figure}[ht]
\centering
\includegraphics[width=\linewidth]{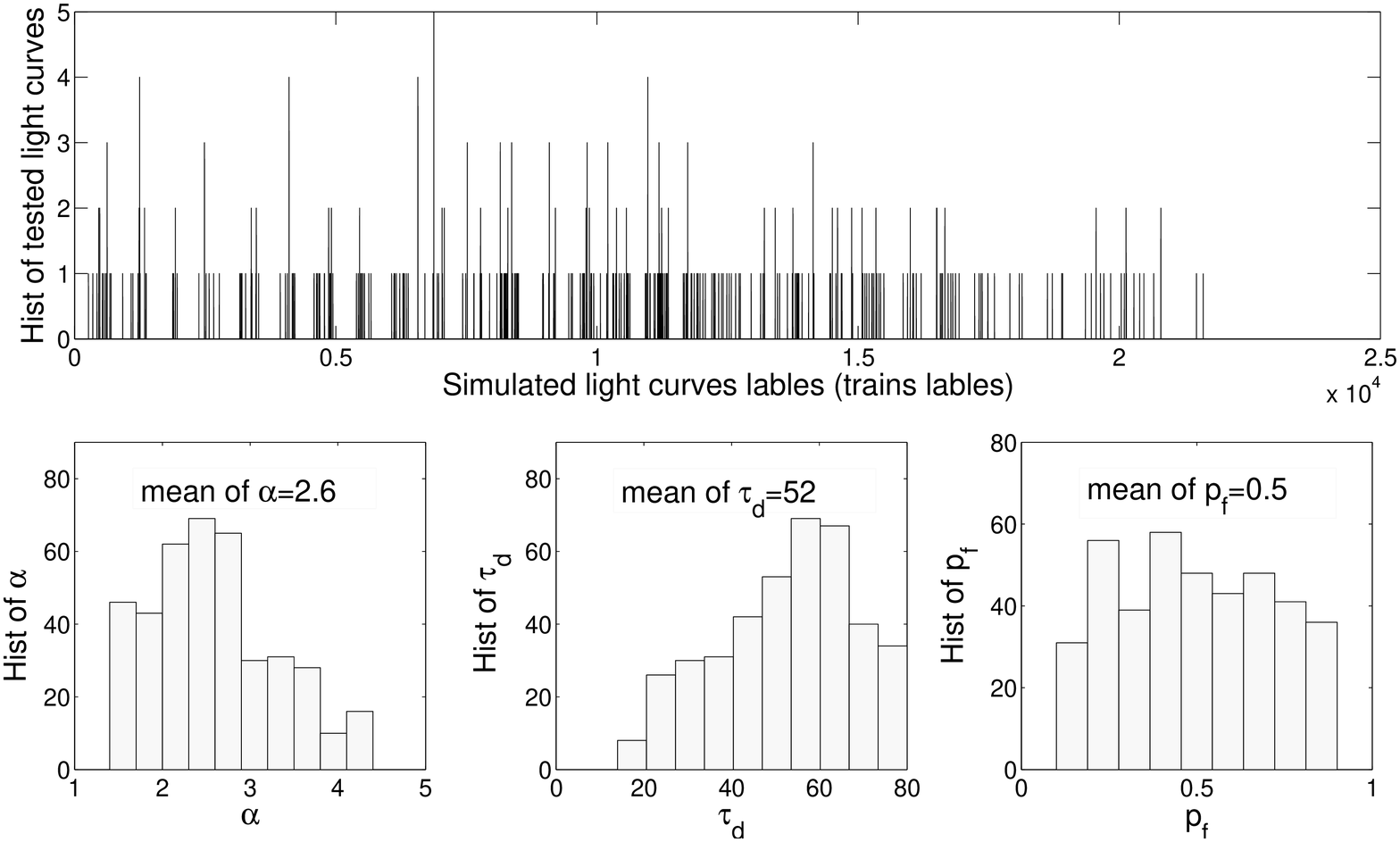}
       \caption{Above panel:  The frequency (histogram) of output labels (simulated light curves labels)
        for the observed light curves (Outside of coronal hole of SDO/AIA image)  constructed by averaging on
        intensities of the smaller region ($3\times3$ pixels). Bottom panel (from left to
right): The histograms of $\alpha$, $\tau_d$, and $p_f$,
respectively.}
            \label{fig12}
\end{figure}

 The outputs
of the classifier network are shown in Figures \ref{fig11} and
\ref{fig12}.  { The average $\alpha$ for both regions are the
same value (i.e., 2.7).  We see that more than 73 percent of
$\alpha$ with $\pm1\sigma_\alpha$ (standard deviation
$\sigma_\alpha=0.70$) falls about mean value.} The average
$\tau_d$ is 48 and 52 for inside and outside the coronal hole,
respectively.  { There are 62 and 64 percent of $\tau_d$ with
$\pm1\sigma_{\tau_d}$ ($\approx$17)  concentrated around mean
values.} This suggests that the background emission in the
coronal hole is lower compared to its surrounding areas. One
point is that there is no considerable difference between
$\tau_d$ inside and outside
 the coronal hole. This is due to the fact that the damping
times should be roughly the same everywhere because by choosing a
specific filter we are picking up the same event phase. The
average  $p_f$ is 0.4 and 0.5 for inside and outside,
respectively.  {Actually, more than 53 and 59 percent of $p_f$s
with $\pm1\sigma_{p_f}$ ($\approx$0.2) are concentrated around
the mean values}.

\begin{figure}[ht]
\centering
\includegraphics[width=\linewidth]{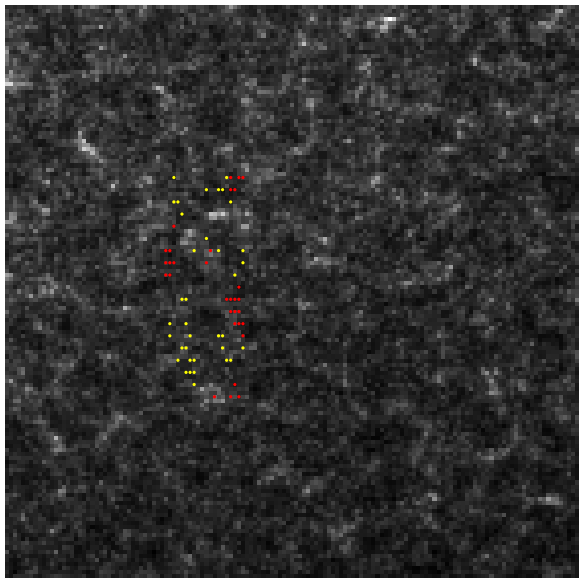}
       \caption{The low $p_f=0.1$ events (red points) fall in the supergranul junctions (dark regions)
         and the high $p_f=0.9$ events (yellow points) fall in the supergranul cell centers  (light regions).
         The image was cropped from the full disk image of SDO/AIA 1600 \AA~ 22 August 2010. }
            \label{fig13}
\end{figure}

In the coronal hole the largest number of events have low flare
rates $p_f\leq$0.2 (Figure \ref{fig11}). This is what one would
expect when there is an almost low background as a consequence of
open magnetic field. Outside the coronal hole, the average flare
rate is $p_f=0.5$ (Figure \ref{fig12}).  It seems that the higher
the $p_f$ is, the smoother the background will be. As shown in
Figure \ref{fig13}, this would explain why the higher $p_f$s are
found in the cell centers (the bright regions of the mentioned
Figure). The greater the variation of the background is, the
lower the $p_f$ value will be. So, this explains why low $p_f$s
are seen at the junctions (the dark regions in Figure
\ref{fig13}).

\begin{figure}[ht]
\centering
\includegraphics[width=\linewidth]{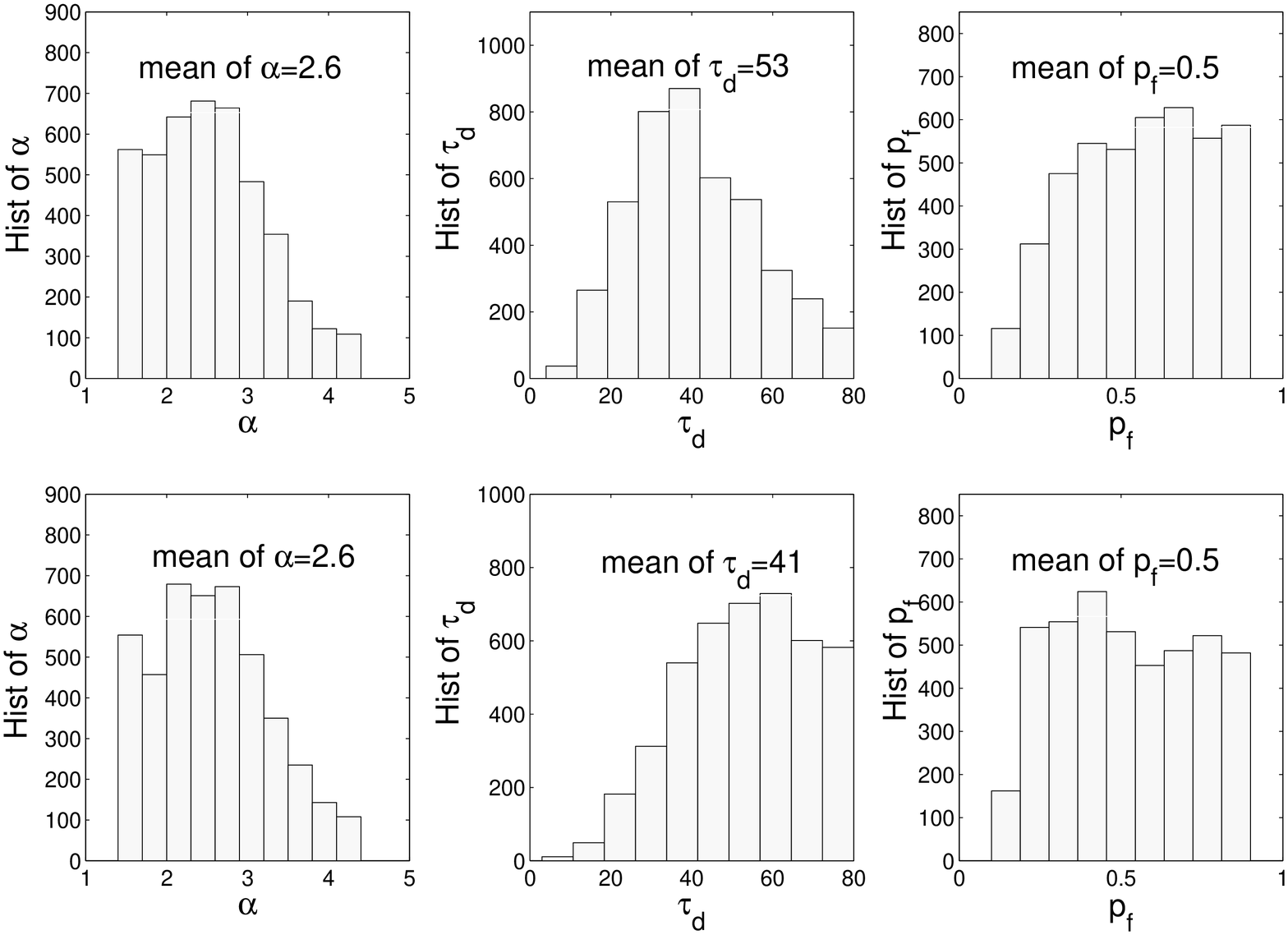}
       \caption{  frequency distribution of
       $\alpha$, $\tau_d$, and $p_f$ for the light curves  of SDO/AIA (Figure 1, test region
       )with cadences of 90s (above panel) and 180s (bottom panel),
       which constructed by averaging on intensities of smaller region ($3\times3$ pixels).}
            \label{fig14}
\end{figure}

\begin{figure}[ht]
\centering
\includegraphics[width=\linewidth]{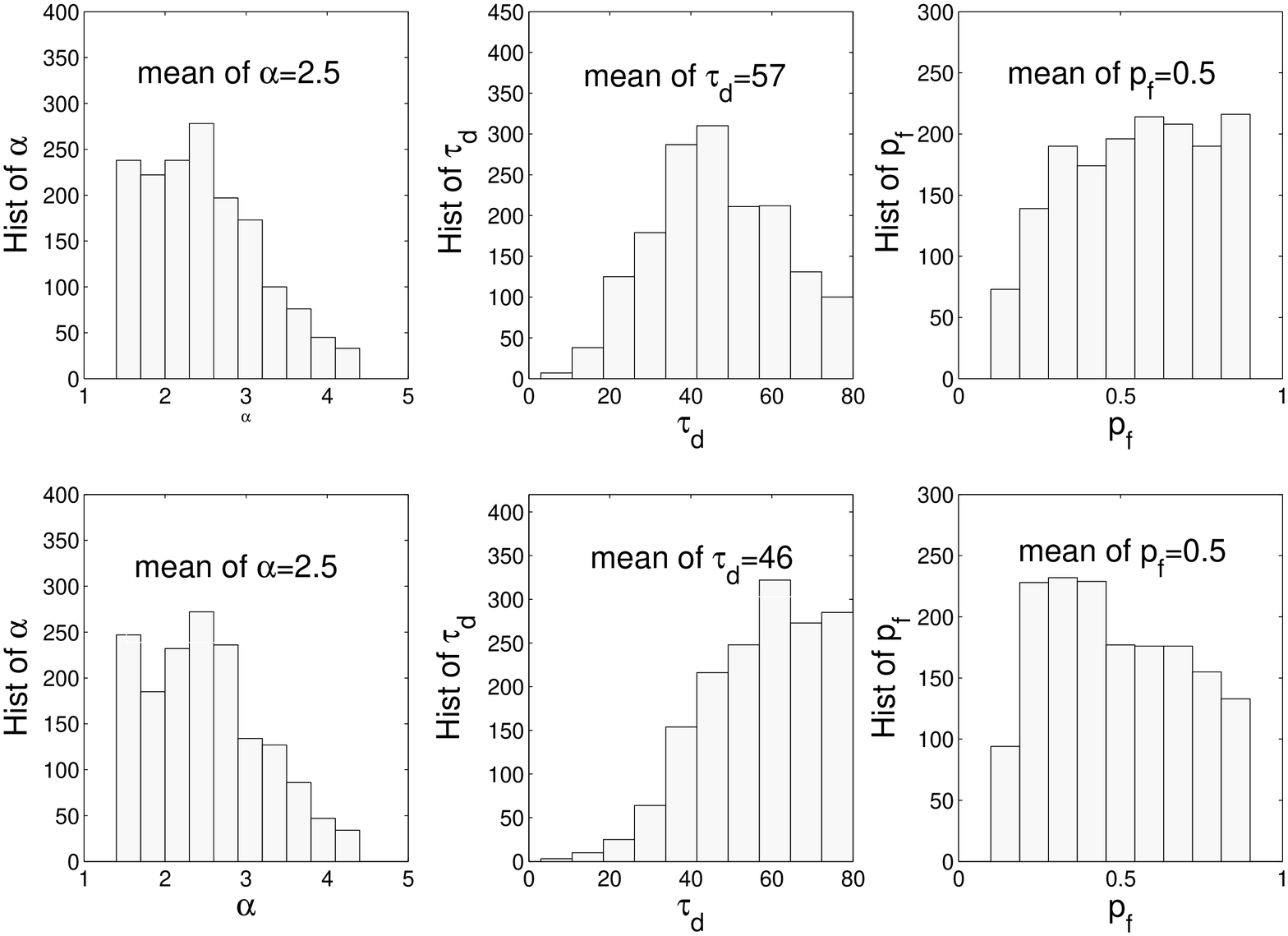}
       \caption{ frequency distribution of
       $\alpha$, $\tau_d$, and $p_f$ for the light curves  of SDO/AIA (Figure 1, test region
       )with cadences of 90s (above panel) and 180s (bottom
       panel), which constructed by averaging on intensities of smaller region ($5\times5$ pixels).}
            \label{fig15}
\end{figure}

The cadence of SDO/AIA (90s for 22 August 2010) is less
than that of STEREO/EUVI (150s for 13 June 2007). As is shown in
Figures \ref{fig8} and \ref{fig12} (we compare this figures
because of their similar averaging dimensions on pixels regions)
the higher the cadence is, the lower the average of damping time
rate (which actually is less than 150/90) will be. For further
examination, we produced AIA light curves  with the cadences of
$90$ and $180$s (for both regions partitioning
 $3\times3$ and $5\times5$ pixels) from the test region in Figure \ref{fig1}.
Again the light curves were fed to the network and the frequency distribution
 of $\alpha$,  $\tau_d$, and $p_f$, were shown in Figures \ref{fig14} and \ref{fig15}.
As we see, there is no change in  average  $\alpha$ and $ p_f$
with changing cadences  but the average  $\tau_d$ decrease while
the cadence is increasing.

\section{Conclusions }\label{conc}

The basic aim of the model, as mentioned earlier, is to
determine the coronal heat input by numerous randomly distributed
small scale events specially nanoflares by finding the power law
index. This will decides  whether the small scale event
contribution is of importance or not. The problem of the too
power law index indicated by  researchers may
have been just a bias due to neglecting the overlapping
nanoflares.

Here, a simple nanoflare model based on three key
parameters (the flare rate, the flare decay time, and the power
law exponent of the flare energy frequency distribution) is used
to simulate emission line radiances from the STEREO/EUVI and
SDO/AIA in the corona. The simulation code ran to generate more
than 22000 light curves (train set) for each combination of
$\alpha$, $\tau_d$, and $p_f$. For each of the marked regions on
the full solar disk images, more than 15000 light curves were
generated for average intensities. This large number of perfect
light curves which enable us in statistical description was focus
of the present-paper.  Light curve pattern recognition by a
Probabilistic Neural Network (PNN) was employed to determine
values of the key parameters. We found that more than 85\% of the
observed light curves have a power law index greater than 2.
Since the network's sensitivity depends on the training set in
which the network must see all the possible patterns during the
training session, we may have some errors in recognizing the
correct patterns.  Empirically, the network is sensitive to steps
 $\Delta \tau_d\approx1$, $\Delta p_f\approx0.1$, and
$\Delta\alpha\approx0.1$ for three key parameters. This means
that, for shorter steps the light curves are too like, so that the
network is not able to classify.

 The results
can be summarized as follows:
\begin{description}
  \item[-] A physical picture of how the model's parameters affect the simulated light curves is
discussed.   Decreases in both  average and variance of the light curves are the
 function of increasing  power law index
   (greater $\alpha$ value corresponds to greater number of small events).
 The higher moments (skewness  and kurtosis) values  of the  time series
are accompanied with increasing $\alpha$, $\tau_d$, and $p_f$
values (because of the lognormal shape of the distributions). Both
the skewness and kurtosis values are  positive numbers.
 {The distributions of both simulated and observed
 light curves are asymmetric (Terzo et al. 2011). }
  \item[-] The average $\alpha$ and $ p_f$ did not change with
  changes in data cadences but the average  $\tau_d$ is a sensitive
function of decrease in cadence.
  \item[-]  With regard to average dimensionless range of $\tau_d$=40-50s and by multiplying it by cadence of $2.5$ min
  (for STEREO/EUVI, 13 June 2007) and 1.5 min (for SDO/AIA, 22 August 2010) we obtain values of
   $100-125$ min and  $100-75$ min, respectively. Assuming that plasma cooling through the narrow
    band filter is dominated by radiative cooling  we find that the ranges are consistent with previous results.
  \item[-] The effect of the background emission on the flare
  rate, $p_f$ is studied. In the coronal hole regions
  with  less background majority  of events has low
flare rates.
\end{description}

The next logical step is to determine the actual flare energies
and the total energy input to the corona which is still a problem
for researchers to be solved in the future.

\acknowledgments The authors thank the unknown referee for
his/her very helpful comments and suggestions

\end{document}